\documentclass[aps,prl,twocolumn,groupedaddress,showpacs]{revtex4-1} %
\usepackage{graphicx}
\usepackage{subfigure}
\usepackage{relsize}
\usepackage{amsmath}
\usepackage{amsfonts}
\usepackage{amssymb}
\usepackage[dvipsnames]{xcolor}
\usepackage{color} 
\usepackage{xcolor}
\usepackage{tikz}
\usetikzlibrary{shapes}
\usepackage{float}
\onecolumngrid
\definecolor{mypink3}{rgb}{0.858, 0.188, 0.478}

\begin{document}
\title{Dissipative light bullets in Kerr cavities: multi-stability, clustering, and rogue waves}
\author{S. S. Gopalakrishnan$^{1,4}$,  K. Panajotov$^{2,3}$, M. Taki$^{4}$, and M. Tlidi$^{1}$}
\affiliation{$^1$ Facult\'{e} des Sciences, Universit\'{e} libre de Bruxelles (U.L.B), CP. 231, 1050 Brussels, Belgium}
\affiliation{$^2$ Department of Applied Physics and Photonics (IR-TONA), Vrije Universiteit Brussels, Pleinlaan 2, 1050 Brussels, Belgium}
\affiliation{$^3$ Institute of Solid State Physics, 72 Tzarigradsko CHaussee Blvd., 1784 Sofia, Bulgaria}
\affiliation{$^4$ Universit\'{e} Lille, CNRS, UMR 8523 - PhLAM - Physique des Lasers, Atomes et Mol\'{e}cules, F-59000 Lille, France}
\date{\today}

\begin{abstract}
We report the existence of stable dissipative light bullets in Kerr cavities. These three-dimensional (3D) localized structures consist of either an isolated light bullet (LB), or could occur in clusters forming well-defined 3D patterns. They can be seen as stationary states in the reference frame moving with the group velocity of light within the cavity. The number of LBs and their distribution in 3D settings are determined by the initial conditions while their maximum peak power remains constant for a fixed value of the system parameters. Their bifurcation diagram allows us to explain this phenomenon as a manifestation of homoclinic snaking for dissipative light bullets. However, when the strength of the injected beam is increased, LBs lose their stability and the cavity field exhibits giant, short living three-dimensional pulses. The statistical characterization of pulse amplitude reveals a long tail probability distribution indicating the occurrence of extreme events; often called rogue waves.
\end{abstract}
 
\maketitle
Confinement of light leading to the formation of localized structures often called
dissipative solitons (DSs), either in space or in time, are realized in cavities with an optical injection (see recent overviews on this issue \cite{akhmediev1997solitons,*Chembo2017theory,*Malomed19,*lugiato_prati_brambilla_2015}). In one-dimensional (1D) dispersive Kerr cavities, temporal \cite{Leo10} and spatial \cite{Taki1} DSs have
been experimentally observed.  The link between the formation of temporal DS
and frequency combs which was established in  \cite{coen2013moctave,*lugiato2018kenberberg}, has reinforced the interest in this field of research. This is because their spectral contents correspond to optical
frequency combs which are sources with a spectrum consisting of millions of equally spaced laser lines. This phenomenon has attracted a lot
of interest since it has applications in the generation of ultra-stable light-wave and microwave
signals which maybe used in optical communication networks, microwave photonic
systems, and in aerospace engineering \cite{ferdous2011spectral,*kippenberg2011microresonator}. 

In two-dimensional (2D) diffractive (broad area) cavities, DSs have been experimentally observed as well  \cite{Taranenko_pra00,*Brambilla3}. They are addressable structures with a possibility for applications in all-optical control of light, optical storage, and
information processing  \cite{Brambilla3}.  Apart from these applications in 1D and 2D settings they are of general interest to scientists through their link to the field of pattern formation such as in hydrodynamics, Bose--Einstein condensates, biology and plant ecology \cite{ROSANOV19961,*Rosanov02,*Akhmediev08,*Mihalache14,*Rev12,*Tlidi5}.

\begin{center}
 \begin{figure}
 \unitlength=40.0mm
\centerline{
\includegraphics[width=3.6\unitlength,height=1.5\unitlength]{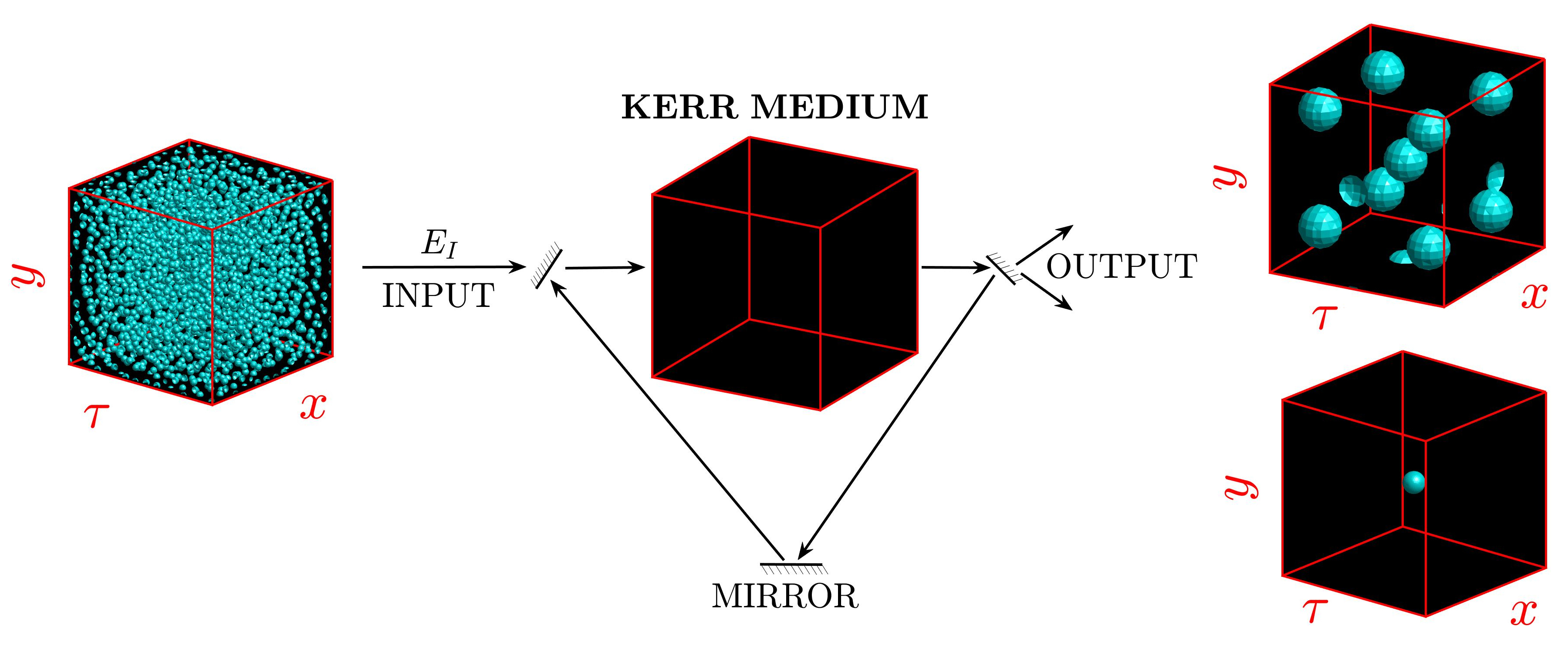}
}
 \caption{Schematic setup of an optical cavity filled with a Kerr medium. The cavity is driven by an external field $E_I$ which evolves a random distribution of optical states as the input. The body-centered-cubic lattice and a single light bullet are shown as possible outputs depending on the seeded initial condition.}
 \label{fig:schem}
\end{figure}
\end{center}

When dispersion and diffraction have the same influence, the formulation of this problem leads to the generalised Lugiato--Lefever equation (LLE,  \cite{Lefever1}) describing the dynamics of nonlinear optical cavities.
An example of such an optical cavity is shown in Fig. \ref{fig:schem}. In its three-dimensional (3D) form the LLE reads \cite{Tlidi2,*Tlidi4}

\begin{equation}
\frac{\partial E}{\partial t} = E_{I} - ( 1 + \dot{\imath}\theta) E + \dot{\imath}\Big(\nabla_{\perp}^2  + \frac{\partial^2}{\partial \tau^2}\Big) E + \dot{\imath}\vert E \vert^2 E.
\label{eqn:LLE}
\end{equation}
Here $E(x, y, \tau, t)$ is the normalized slowly-varying envelope of the intracavity field, $E_I$ is the input field, and $\theta$ is the detuning parameter. The 2D diffraction  is described by the Laplace operator $\nabla_{\perp}^2= \partial_{xx} + \partial{yy}$ acting on the transverse plane $(x, y)$. 
The time $t$ corresponds to the slow-time evolution of $E$ over successive round-trips, whereas $\tau$ accounts for the fast time in a reference frame traveling at the group velocity of light in the Kerr medium. We consider the case of anomalous dispersion regime as indicated by the $+$ sign in front of the second derivative with respect to $\tau$.
 
The purpose of this Letter is to use Eq. (\ref{eqn:LLE}) to predict the occurrence of stable stationary light bullets (LBs), that can be either isolated or bounded together as self-organized clusters. Some of these 3D structures have been reported in the limiting case of nascent bistability in non-Kerr media \cite{Staliunas1,*Tlidi3}. 
Here we characterise these 3D solutions in Kerr media by constructing their bifurcation diagram and determine their range of stability. This diagram exhibits multi-stability, which clearly indicates that clustering phenomenon belongs to a homoclinic snaking type of bifurcation. When increasing the injected field power, we observe transition via period doubling to a complex dynamical regime. In this regime, numerical analyses shows that the amplitude probability distribution possesses a long tail with pulse intensity height well beyond two times the significant wave height \cite{kharif2008rogue}. From these two characteristics we can infer that this complex behavior belongs to the class of rogue waves. Bounded states of LBs and rogue waves in three-dimensional settings have neither been experimentally observed nor theoretically predicted. Temporal solitons consisting of 2$\pi$ phase rotation in spatially multimode ring cavity semiconductor lasers has been recently experimentally established \cite{gustave2017observation}. We address here the theoretical side for the realization of dissipative optical bullets.

\onecolumngrid
\twocolumngrid

The homogeneous stationary solutions of LLE, $E_S$, are given by 
$E_{I}^{2} = \vert E_S \vert^{2}[1 + (\theta - \vert E_S \vert^{2})^{2}]$. For $\theta < \sqrt{3}$, the transmitted intensity as a function of the input intensity $E_{I}^{2}$ is single-valued, whereas bistability occurs for $\theta > \sqrt{3}$. The steady-state homogeneous solution undergoes a Turing instability when the input field intensity is raised above the threshold value, $ E_{c}^{2} = 1 + (\theta - 1)^2$, corresponding to the intracavity intensity $\vert E_S \vert^{2} = 1$, beyond which they become unstable \cite{Lefever1}. At this bifurcation point, the wavelength of 3D patterns is $\Lambda=2\pi/\sqrt{2-\theta}$.
A weakly nonlinear theory has been performed in 1D, 2D, and 3D settings \cite{Tlidi2}. This analysis allows one to determine the variety of 3D periodic structures which are solutions of Eq. (\ref{eqn:LLE}) in the
weakly nonlinear regime where the Turing bifurcation, often called modulational instability, is
supercritical. This analysis has revealed the predominance of the body-centered-cubic lattice structure (BCC) over other three-dimensional periodic structures such as face-centered-cubic or hexagonally packed cylinders. 
This analysis through the normal mode theory cannot explain 
the formation of LBs since it does not take into account of the fast spatial scales. Moreover, this weakly nonlinear analysis 
is restricted to the values of the detuning parameter in the range $\theta<41/30$. 

In what follows we focus on the strongly nonlinear monostable regime where
modulational instability is subcritical, i.e., $41/30<\theta<2$. Remarkably, besides the emergence of  BCC structures, the same mechanism predicts the possible existence of aperiodic distribution of dissipative light bullets. An example of a single light bullet is shown in Fig. \ref{fig:schem} together with the BCC periodic structures. These structures are obtained by a direct numerical simulation of Eq. (\ref{eqn:LLE}) for the same values of parameter and by using periodic boundary conditions in all directions. They differ only by the initial condition used. The BCC structure can be obtained by a small amplitude noise as the seeded initial condition. However for the formation of a single LB, a localized Gaussian shell is used given by $E_0 = E_s + \exp\big[-10\big({(x - x_0)^2}+(y - y_0)^2 +(\tau - \tau_0)\big)/L\big]$, where $L$ is the length of the periodic domain, with $(x_0, y_0, \tau_0)$ the center of the localised Gaussian shell. 
The cross-section of the LB is shown in Fig. \ref{fig:Fig2_1LB_Summary}(c) 
along the transverse plane, with the spatial profile indicating that LBs possess a damped oscillatory tail.  
This oscillatory tail connects the LB with the underlying BCC lattice.
Stationary, localized, and spherically symmetric solutions of LLE (\ref{eqn:LLE}) 
have the form $E(r) = E_s(1 + A(r))$ with boundary conditions 
$A(0) = A_0 (\neq 0)$, $\partial_r A(0) = 0$, and $A(r \rightarrow \infty) = 0$. 
This is shown using a dotted red line in panel (b), with the numerically converged solution plotted in black. The width of LB is close to $\approx\pi/\sqrt{2-\theta}$ which is half the critical wavelength at the modulational instability.

\begin{center}
 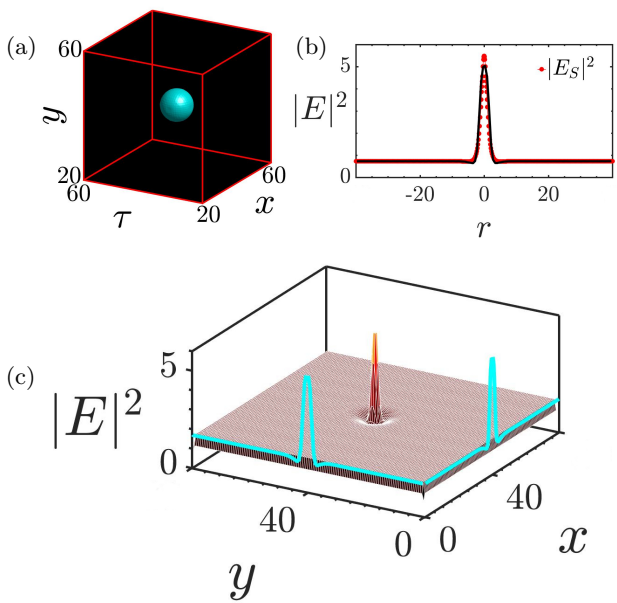
\begin{figure}
 \unitlength=13.0mm
 \centerline{
 	\includegraphics[width=2.5\unitlength,height=2.5\unitlength]{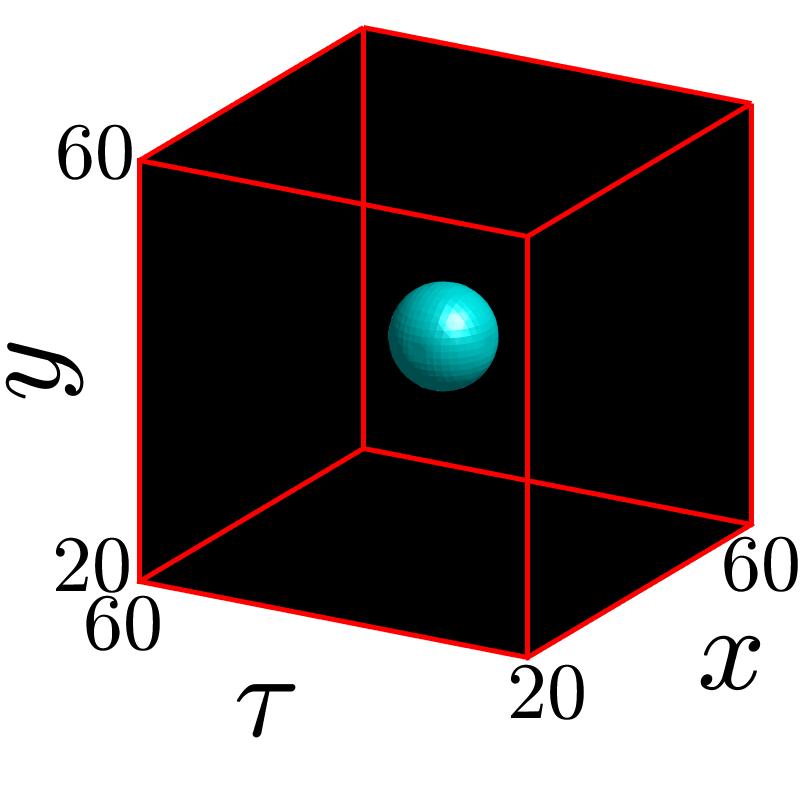}
 	\includegraphics[width=3.375\unitlength,height=2.25\unitlength]{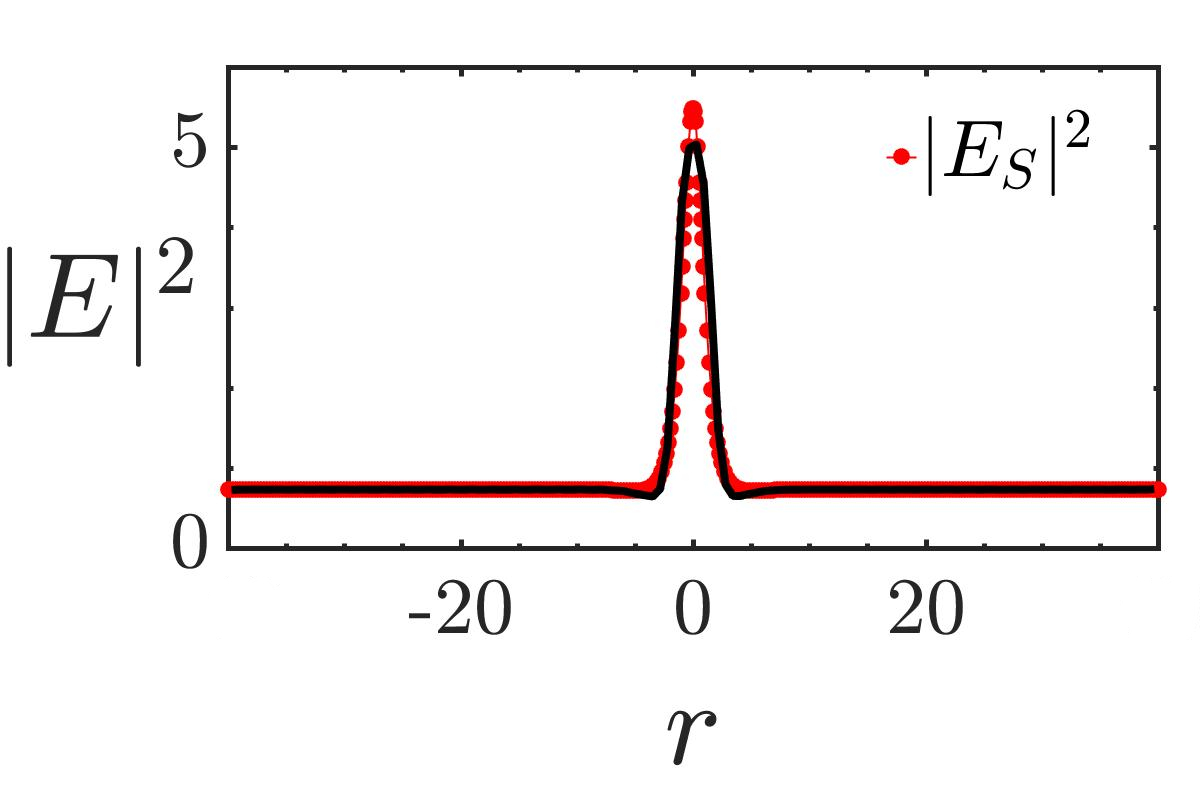}
 }
  \unitlength=22.0mm
\centerline{
\includegraphics[width=3.375\unitlength,height=2.25\unitlength]{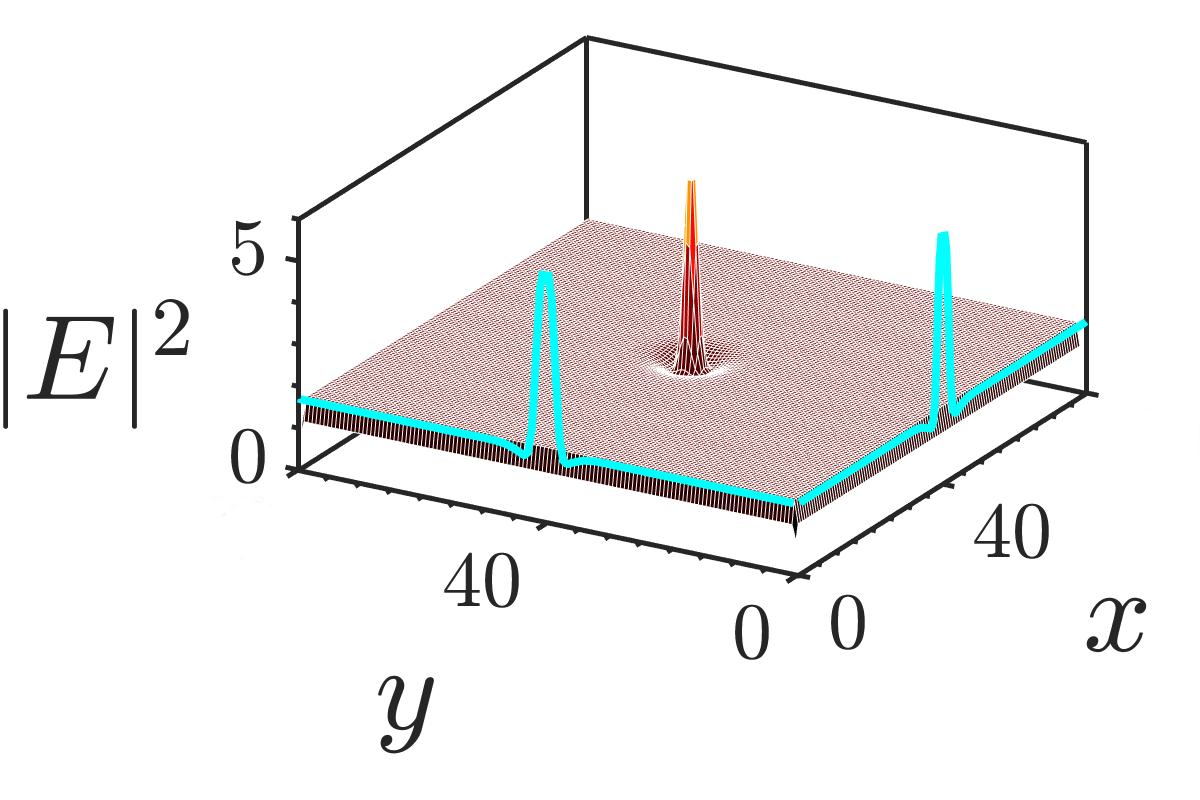}
}
 \caption{\small{A single robust light bullet. 
 (a) Stable 3D LB obtained by a direct numerical simulation of Eq. (\ref{eqn:LLE}). The 
 isosurface at $1\%$ of the maximal value is shown. (b) The one-dimensional perspective of the LB along with the 
steady-state solution obtained from spherical symmetry considerations which is shown using a red dotted line.
 (c) The cross-section along the transverse $(x,y)$ plane of the LB with the profiles exhibiting a decaying spatial oscillation. 
 Parameter settings: $E_I = 1.21, \theta = 1.7$. }
}
\begin{picture}(0,0)
\put(-1.95,3.3){(c)}
\put(-1.95,5.3){(a)}
\put(-0.2,5.3){(b)}
\end{picture}
 \label{fig:Fig2_1LB_Summary}
\end{figure}
\end{center}
\vspace{-1.0cm}
\normalsize

From a theoretical point of view the challenging aspect lies in the instability and collapse of DSs for the case of the nonlinear Schr\"{o}dinger equation when the dimensionality of the system is at least two \cite{Silberberg1, *Edmundson1}. Wave collapse is accompanied by the formation of singularities in the wave packet within a finite time. To avoid this phenomenon, it is necessary to introduce additional physical effects, such as semiconductor active media \cite{Brambilla1, *Brambilla2}, saturable absorbers \cite{Rozanov1,*Veretenov1,*Marconi1,*Javaloyes1,*Gurev1}, parametric oscillators \cite{Staliunas1,*Veretenov2}, etc. (for recent reviews see \cite{Malomed1,*Mihalache14,*Malomed19}). Numerically, the challenge posed by the 3D LLE arises from the strongly nonlinear term, which when discretized leads to large systems of strongly nonlinear stiff ordinary differential equations \cite{Jones1,*Kassam2}. 
Due to this reason the 3D simulations of LLE are still missing in the literature.
It is to be noted that finite-difference methods can sometimes lead to spurious solutions which are non-physical \cite{Jones1}, which is where higher-order spectral methods come to the fore.
In the present work, the spatial discretization of the LLE is done using a Fourier spectral method with periodic boundary conditions \cite{Trefethen1, *Kassam2,*Saad1}, and the time-stepping is carried out with a fourth order exponential time differencing Runge--Kutta method \cite{Trefethen1, *Matthews1}. The main advantage of using a Fourier spectral method lies in the fact that the linear term of the discretized set of equations is diagonal, and more importantly the nonlinear term is evaluated in physical space and then transformed to Fourier space. Further details on these methods can be found in these excellent books \cite{Trefethen1, *Kassam2,*Saad1}. All the numerical simulations in the present study are carried out on a periodic domain of size $80$ units in each direction 
resolved using $128$ grid points, with a time-step of $0.01$.

\begin{figure}
 \unitlength=18.0mm
\centerline{
\includegraphics[width=1.5\unitlength,height=1.5\unitlength]{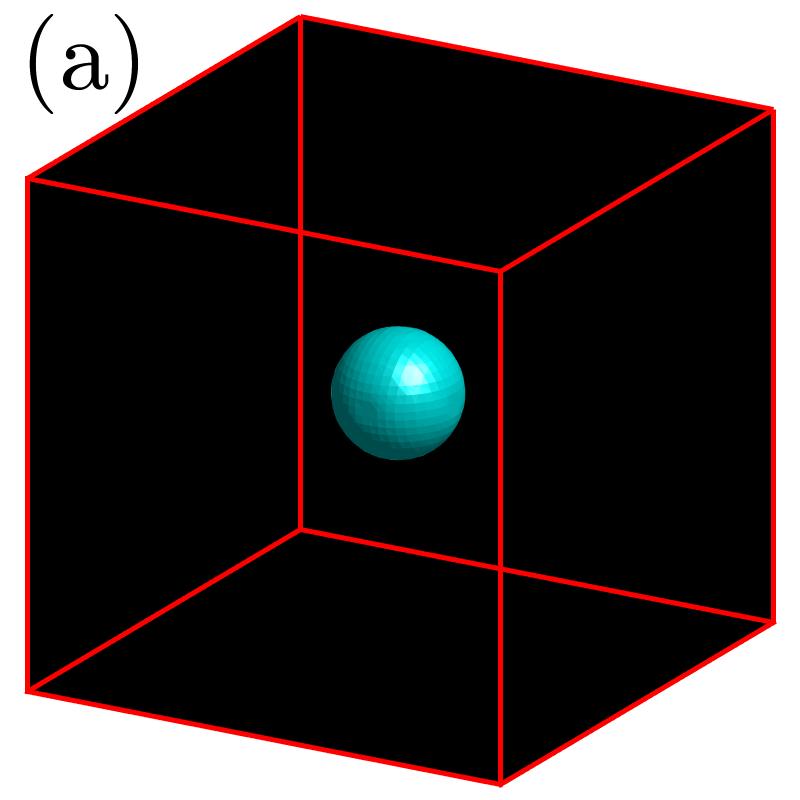}
\includegraphics[width=1.5\unitlength,height=1.5\unitlength]{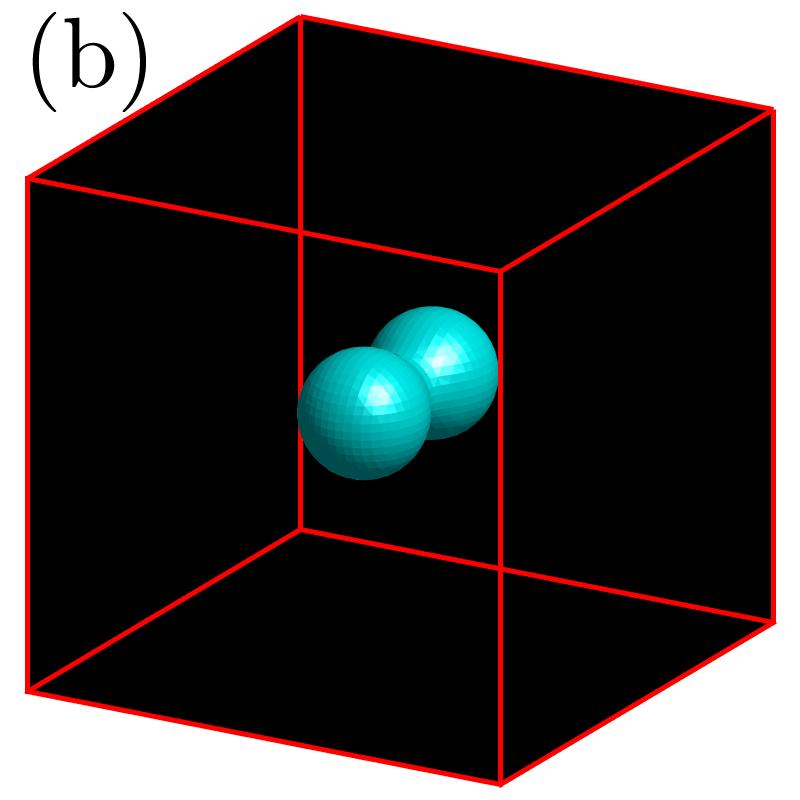}
\includegraphics[width=1.5\unitlength,height=1.5\unitlength]{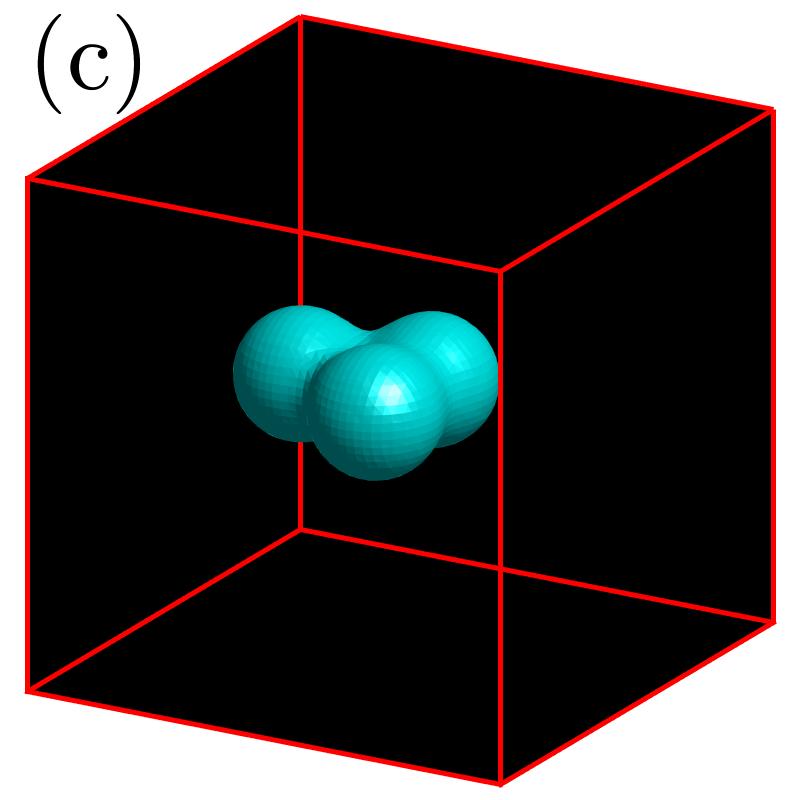}
}
\centerline{
\includegraphics[width=1.5\unitlength,height=1.5\unitlength]{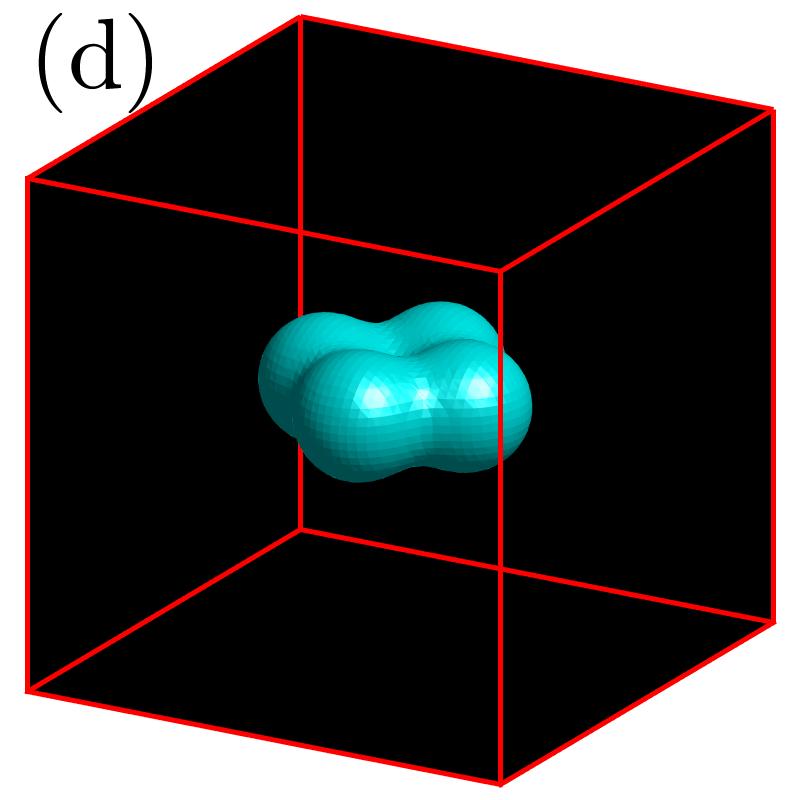}
\includegraphics[width=1.5\unitlength,height=1.5\unitlength]{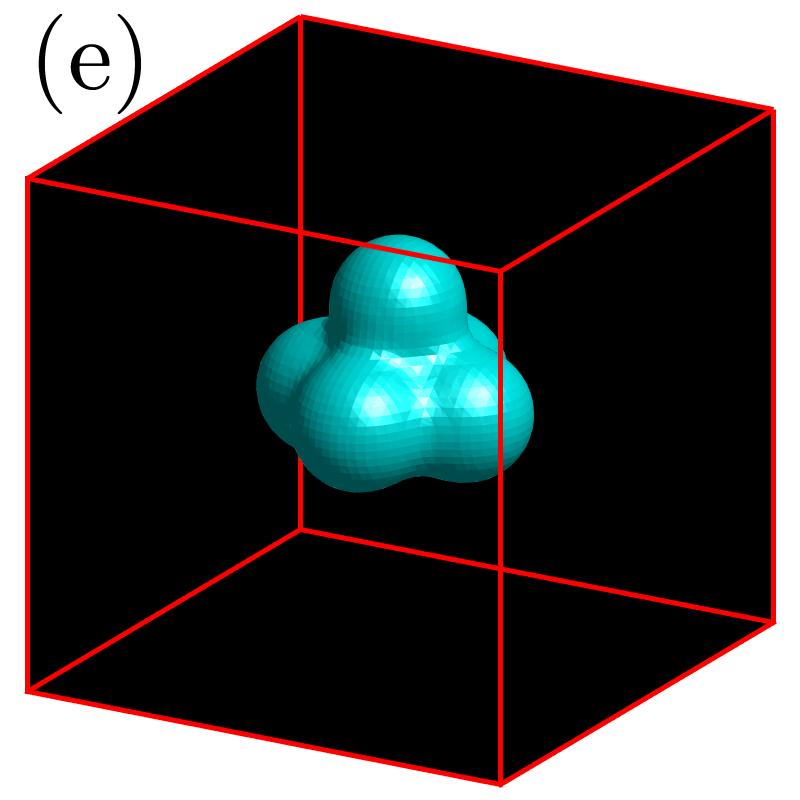}
\includegraphics[width=1.5\unitlength,height=1.5\unitlength]{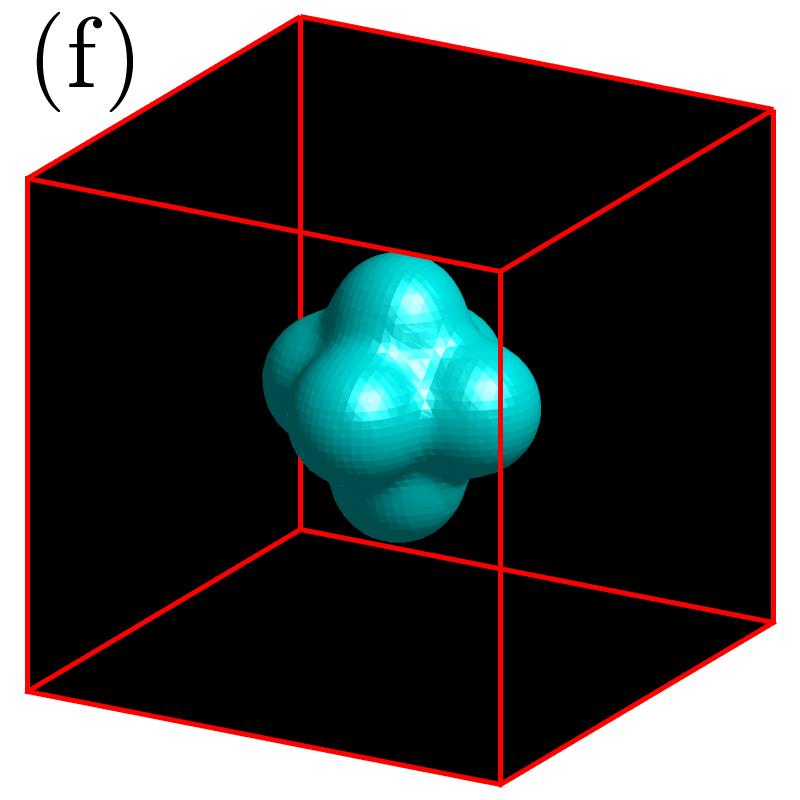}
}
\vspace{0.6cm}
 \unitlength=18.0mm
 \centerline{
\hspace{-0.5cm}\includegraphics[width=3.8\unitlength,height=3.8\unitlength]{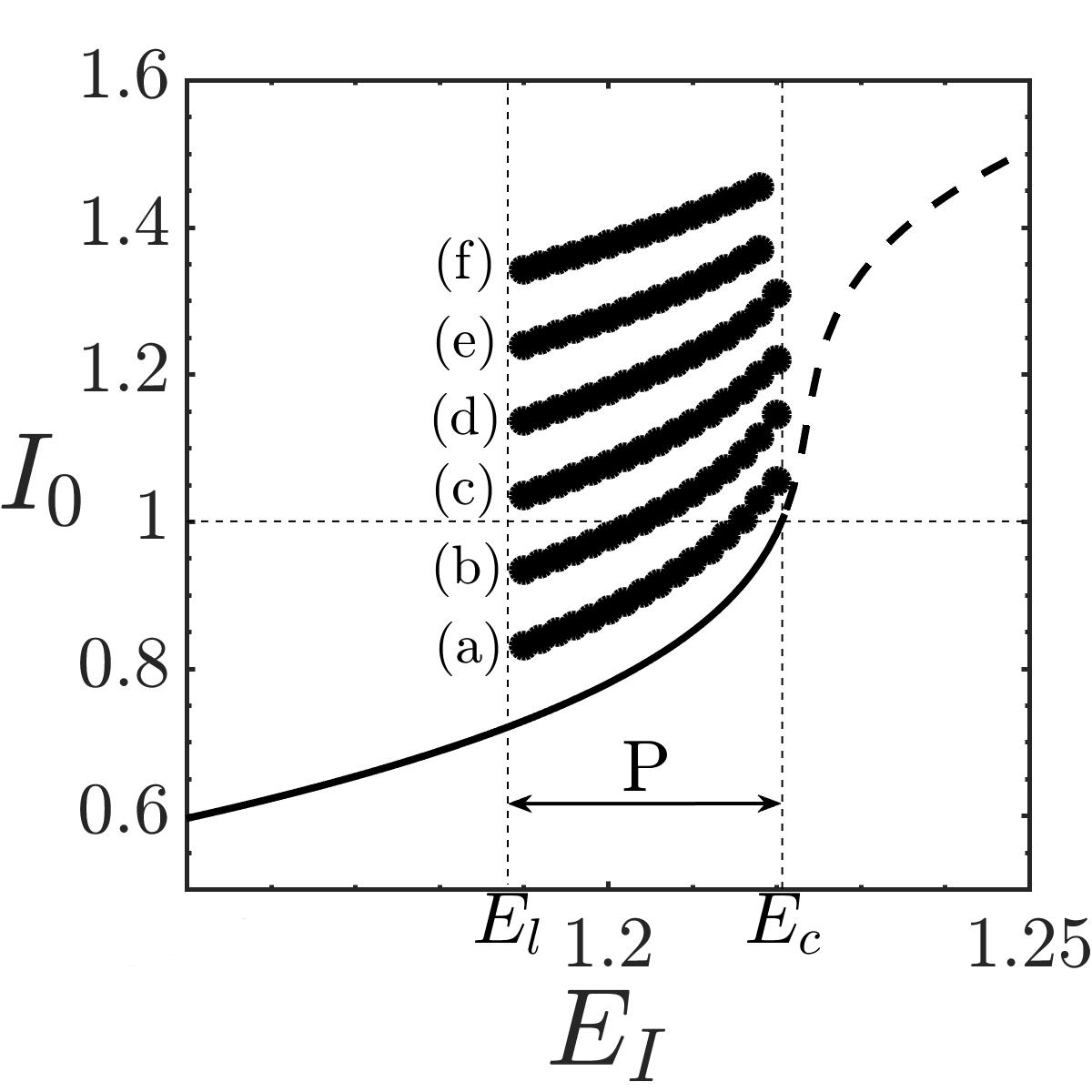}
}
\begin{picture}(0,0)
\put(-1.15,3.78){\includegraphics[width=0.6\unitlength,height=0.7\unitlength]{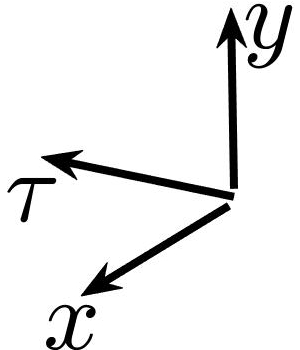}}
\end{picture}
\vspace{-0.6cm}
\caption{\small{Bifurcation diagram for the various clusters of closely packed bounded states. The LBs are characeterised using $I_0 = \int\vert E - E_S \vert^2 dx\,dy\,d\tau$. The continuous black line denotes the stationary steady-state, with the critical value of $E_I$ being $E_c = 1.22$ ($\theta = 1.7$). The pinning range is indicated by $\textrm{P}$.
(a,b,c,d,e,f) Different clusters of bounded states, which are robust and localized, form a bifurcation diagram, with the labels indicating the different clusters of closely packed LBs.}
}
 \label{fig:BD}
 \end{figure} 
\normalsize

When initial conditions are set to two Gaussian shells centered at different positions in $E(x, y, \tau)$ space, these two perturbations evolve towards the formation of two bounded LBs as can be seen on Fig. \ref{fig:BD} (b).  By placing different LBs close to each other in the cavity, clusters of stable bounded states of LBs can be formed. When the individual LBs are well separated from one another, they are independent. However, when they are brought closer to each other they start to interact via their 
oscillatory, exponentially decaying tails. The number of LBs and their distribution in the $(x, y, \tau)$ space is large.
Some examples of closely packed LBs are shown in Fig. \ref{fig:BD}(c,d,e,f). 
These solutions are obtained for the same values of parameters, and they differ only by the seeded initial conditions. 
The interaction between the LBs then leads to the formation of closely packed clusters.
All these 3D localized dissipative LBs co-exist as stable solutions with the BCC lattice in the range $\textrm{P}$ shown in Fig. \ref{fig:BD}, often called the pinning range \cite{POMEAU19863}. 
These 3D LBs are confined in $(x, y, \tau)$, and can be seen as a cluster of the elemental structure (a single LB) with a well-defined size, which are robust and stable. However, in contrast to BCC, they apparently have no tendency to spread and invade the whole space available in the 3D cavity. 

\begin{figure}
 \unitlength=22.0mm
\centerline{
\includegraphics[width=1.5\unitlength,height=1.5\unitlength]{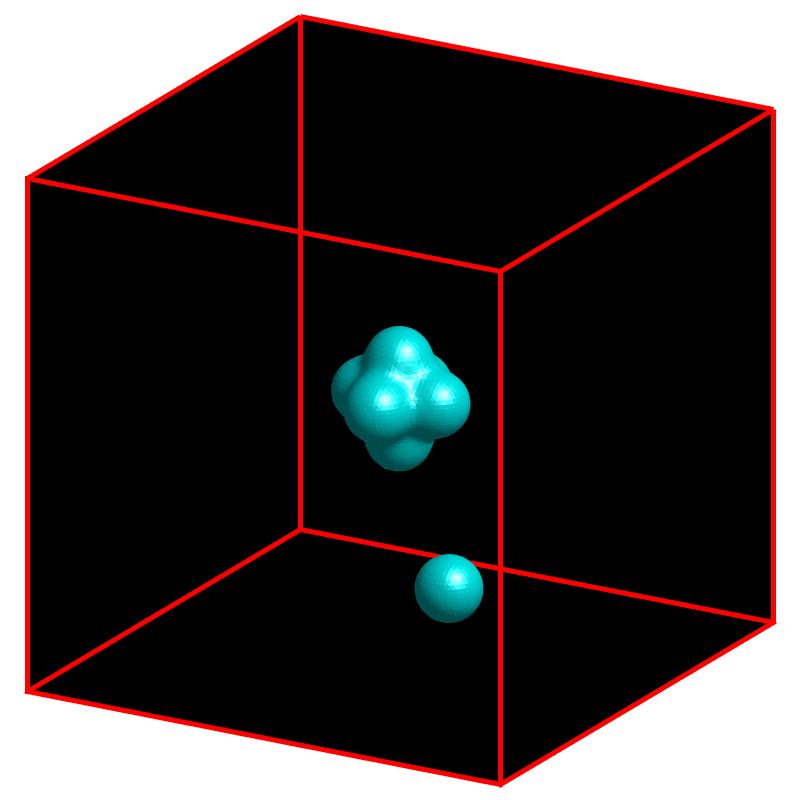}\hspace{0.375cm}
\includegraphics[width=1.5\unitlength,height=1.5\unitlength]{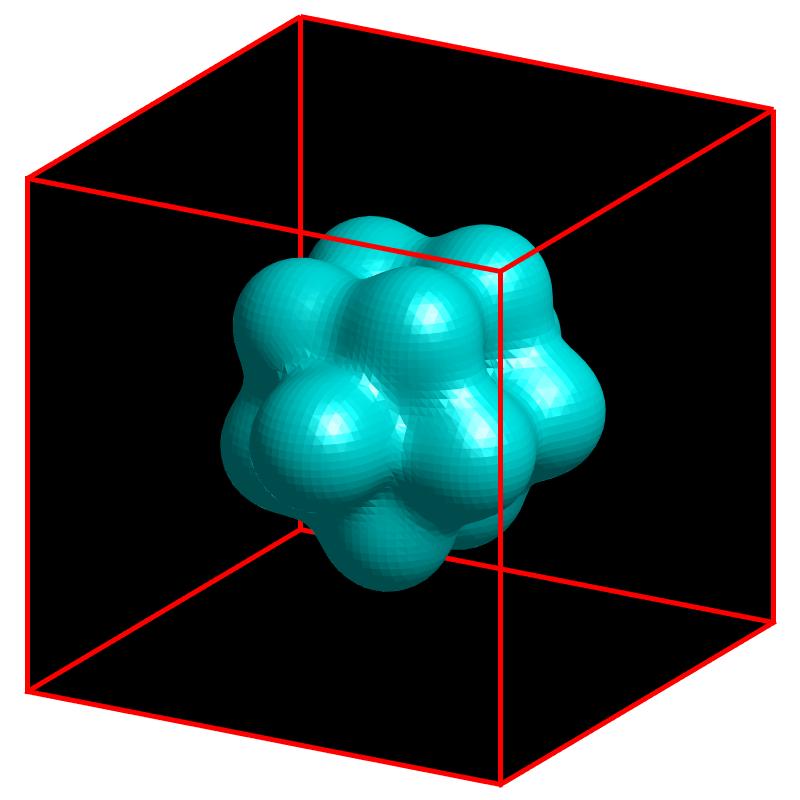}
}
\begin{picture}(0,0)
\put(-1.6,1.55){(a)}
\put(0.1,1.55){(b)}
\end{picture}
 \unitlength=18.0mm
\begin{picture}(0,0)
\put(-0.435,-0.25){\includegraphics[width=0.6\unitlength,height=0.7\unitlength]{Figures/Fig3_Axes_coord.jpg}}
\end{picture}
\vspace{0.25cm}
 \caption{\small{Other configurations of closely packed LBs for the same parameter settings as in Fig. \ref{fig:BD}.
 (a) A cluster of LBs co-existing with an isolated robust LB. In (b) the cluster of LBs correspond to a combination of structures (c) and (f) shown in  Fig. \ref{fig:BD}(c,f).}
}
 \label{fig:BS}
 \end{figure}

The bifurcation diagram of these closely packed clusters of LBs is shown in Fig. \ref{fig:BD} where the difference in intensity of the LB with respect to the background state, $I_0 = \int\vert E - E_S \vert^2 dx\,dy\,d\tau$ is shown as a function of the injected field $E_I$ for a fixed value of the detuning parameter $\theta$. 
The associated label is denoted next to the bifurcation curves. In the pinning range $\textrm{P}$ the system exhibits high degree of multi-stability.
 The multiplicity of these 3D solutions of the LLE is strongly reminiscent of homoclinic snaking indicating that the formation of bounded LB states and clusters of them is indeed a very robust phenomenon. Homoclinic snaking type of bifurcation has been investigated in one dimensional settings  \cite{gomila2007bifurcation,*parra2014dynamics}. This phenomenon however has never been documented in 3D settings. It was first studied in relation with Swift--Hohenberg equation \cite{woods_heteroclinic_1999,*coullet2000stable}, and is also a well documented issue in spatially extended systems (see overviews \cite{burke2007homoclinic,*knobloch2015spatial}). Besides the closely packed LBs shown in Fig. \ref{fig:BD}(b,c,d,e,f), a large number of other configurations of multiple clusters are stable within the domain $\textrm{P}$. Examples of such configurations are shown in Fig. \ref{fig:BS}(a,b). These clusters of 
closely packed LBs are found as isolated elements, or may form groups which gathers a small number of nearby elements.

\normalsize

\begin{center}
 \begin{figure}
 \unitlength=14.0mm
 \centerline{
\includegraphics[width=2.25\unitlength,height=2.25\unitlength]{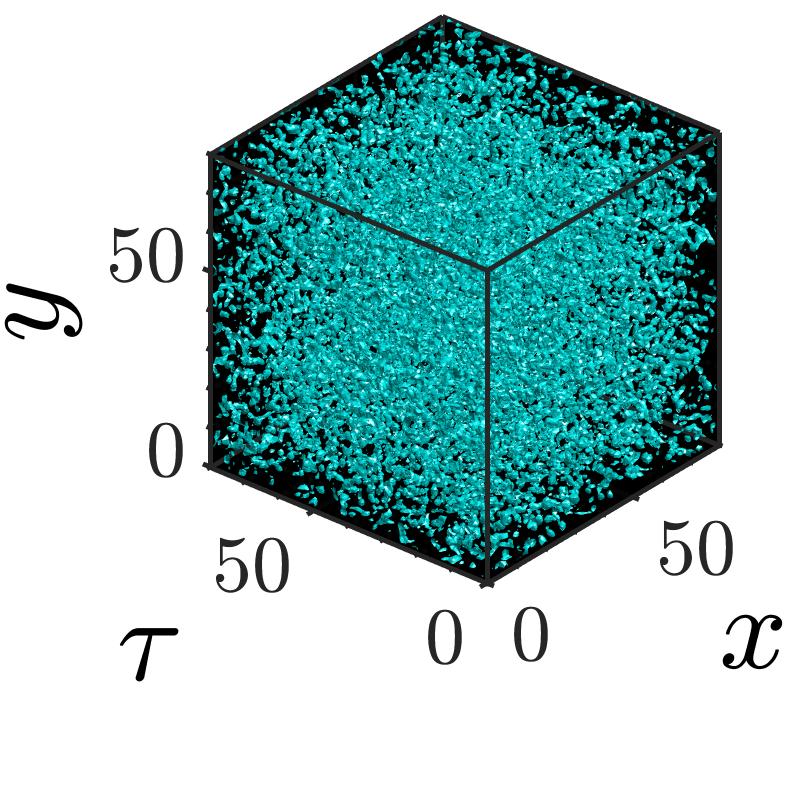}
 \includegraphics[width=3.25\unitlength,height=2.25\unitlength]{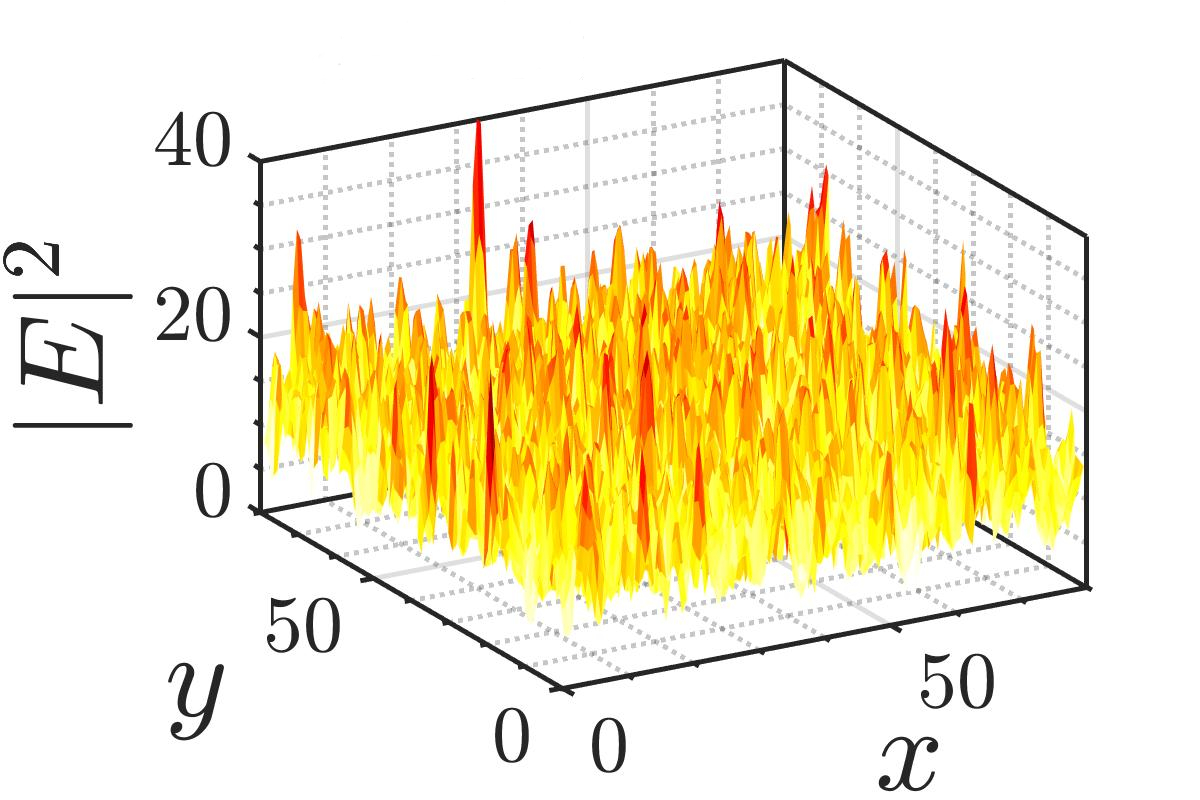}
}
 \centerline{
\includegraphics[width=4.8\unitlength,height=3.0\unitlength]{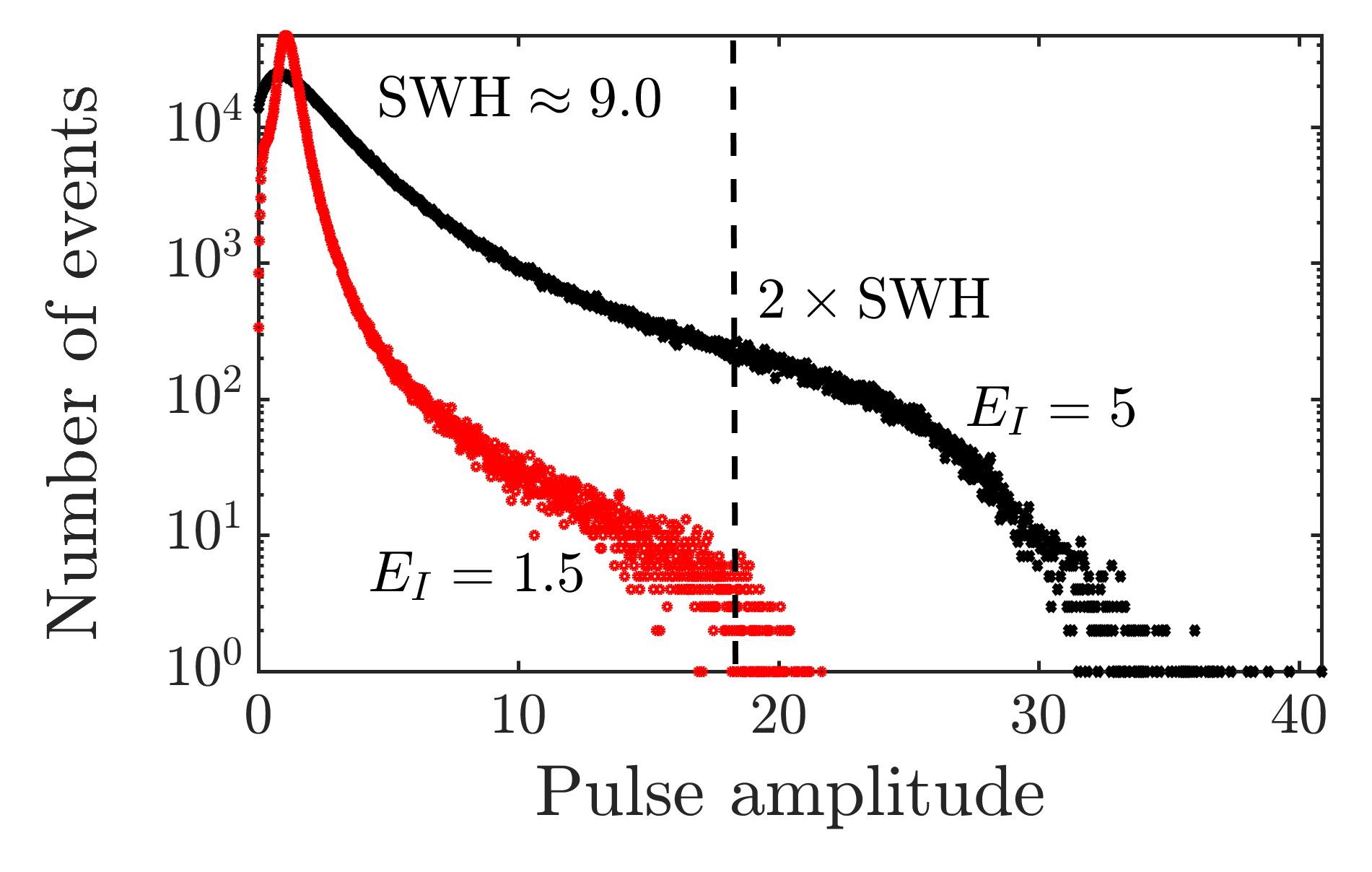}
}
\begin{picture}(0,0)
\put(-2.6,5.25){(a)}
\put(-0.5,5.25){(b)}
\put(-2.6,2.85){(c)}
\end{picture}
\vspace{-0.5cm}
 \caption{\small{Three-dimensional rogue waves.
 (a) A 3D isosurface of rogue waves in the cavity at an input field above $E_c$ ($E_I = 5$) in a complex regime. 
 (b) The cross-section along the transverse $(x,y)$ plane reveals an extreme event.
 (c) The probability distribution of the number of events as a function of the intensity of the pulses in semi-logarithmic scale for two different 
 values of $E_I$ above $E_c$. The dashed line indicates events of amplitudes twice the SWH for $E_I=5$. Parameter settings: $\theta = 1.7$.}
}
 \label{fig:RW_Summary}
\end{figure}
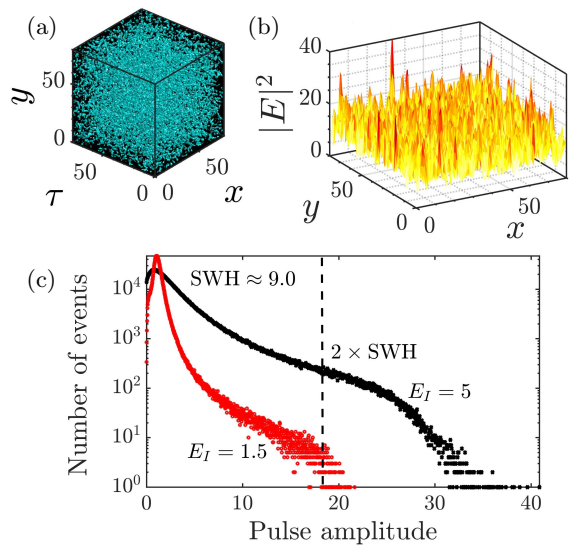
\end{center}
\normalsize
When increasing the injected field intensity $E_I$ beyond the pinning range $\textrm{P}$, LBs are observed to undergo transition via period-doubling to a complex spatio-temporal regime. In the remainder of this Letter, we will characterize this complex regime. At high values of $E_I$ the occurrence of rogue waves in
 one- \cite{coillet2014optical} and in two-dimensional optical Kerr cavities has been recently shown \cite{Tlidi8,*Tlidi1}. The research on extreme events or rogue waves has gathered significant interest, especially in the fields of hydrodynamics, and nonlinear optics \cite{Haver1,*Solli1,*Akhmediev2,*Onorato1,*Akhmediev3,*McAllister1,*KrassimirZhang2020}. 
Rogue waves are rare pulses with amplitudes significantly larger than the average one which arise due to self-focusing of energy of a wave group into one majestic event. We investigate this phenomenon in 3D settings, in the monostable regime.

Fig. \ref{fig:RW_Summary}(a) shows a 3D isosurface of the intensity in the cavity.  Fig. \ref{fig:RW_Summary}(b) shows a two-dimensional cut along the transverse plane demonstrating an event with optical intensity significantly 
larger than the average amplitude of the background state. The three-dimensional perspective [Fig. \ref{fig:RW_Summary}(a)] shows the complex spatiotemporal regime at this high value of $E_I$.
To quantify such extreme events, starting from a random initial condition a statistical analysis based on the pulse amplitudes observed within the optical cavity is registered. 
Fig. \ref{fig:RW_Summary}(c) shows the probability distribution of the number of events as a function of the intensity of the pulses in 
semi-logarithmic scale for two different values of $E_I$. When $E_I$ is increased beyond $E_c$, though the system enters into a complex dynamical regime, the 
tail of the pulse height distribution stays below the threshold of $2$ $\times$ SWH. This can be seen in Fig. \ref{fig:RW_Summary}(c) where the statistical 
analysis at $E_I=1.5$ is shown in red. As $E_I$ is further increased, there is a considerable
number of events with the maxima of
intracavity intensity more than twice the significant wave
height (SWH), with even events of pulse amplitude as high as
$4$ $\times$ the SWH. This can be seen from the statistical analysis shown in Fig. \ref{fig:RW_Summary}(c) for $E_I=5$. 
In this complex spatiotemporal regime the non-Gaussian statistical distribution of the wave intensity can be clearly 
seen with a long tail probability distribution. As remarked earlier, this is the main signature typical of rogue wave formation. 
These large intensity pulses belong to the class of rogue waves or extreme events \cite{Akhmediev3,*Tlidi1}. 
We would like to emphasize that
rogue waves are only formed in the LLE model when the spatiotemporal complexity is well developed, i.e. when the neighboring pulses in the oscillating pattern are interacting strongly. Even when the peak amplitude of the pulses clearly display complex dynamics, as in the case of $E_I = 1.5$
[fig. \ref{fig:RW_Summary}(c)], no rogue waves are formed in the system. This is reflected in the tail of the pulse height distribution which stays below the threshold of 2 $\times$ SWH.

In conclusion, we have explicitly demonstrated the existence of robust three-dimensional dissipative solitons in the form of light bullets leading to a striking light power confinement. The LBs result from the combined action of dispersion and diffraction mediated by Kerr nonlinearity, dissipation and pumping.
They are isolated light bullets, or clusters of bounded states. These localised structures co-exist with the three-dimensional periodic BCC lattice in optical cavities with instantaneous Kerr nonlinearity. They are dissipative three-dimensional structures that travel with the group velocity of light within the Kerr cavity.
These individually addressable LBs can bring a significant benefit for 3D optical storage paving the way for future experimental research. 
When increasing the intensity of the injected field, we have observed the transition from a stationary distribution of LBs to a complex regime. The statistical analysis has revealed evidences of three-dimensional rogue waves with abnormal amplitudes larger than twice the significant wave height.

This work was supported in part by the Fonds Wetenschappelijk Onderzoek-Vlaanderen FWO (G0E5819N). We also acknowledge the support from the French National Research Agency (LABEX CEMPI, Grant No. ANR-11- LABX-0007) as well as the French Ministry of Higher Education and Research, Hauts de France council and European Regional Development Fund (ERDF) through the Contrat de Projets Etat-Region (CPER Photonics forSociety P4S). This work was supported in part by the Fonds Wetenschappelijk Onderzoek-Vlaanderen FWO (G0E5819N). M. Tlidi is  a Research Director at the Fonds National de la Recherche Scientifique (Belgium).

\bibliography{BiblioLLE3D}

\end{document}